\documentclass[sigconf,screen,]{acmart}

\AtBeginDocument{%
  }

\acmYear{2025}\copyrightyear{2025}
\setcopyright{rightsretained}
\acmConference[FSE '25]{33rd ACM International Conference on the Foundations of Software Engineering}{June 23--28, 2025}{Trondheim, Norway}
\acmBooktitle{33rd ACM International Conference on the Foundations of Software Engineering (FSE '25), June 23--28, 2025, Trondheim, Norway}
\acmDOI{10.1145/3696630.3728717}
\acmISBN{979-8-4007-1276-0/25/06}

\usepackage{natbib}

\usepackage{amsmath,amssymb,amsfonts}
\usepackage{algorithmic}
\usepackage{rotating}
\usepackage{array}
\usepackage{graphicx}
\graphicspath{{figures/}}
\usepackage{textcomp}
\usepackage{xcolor}
\usepackage{blindtext}
\usepackage{balance}
\usepackage{multirow}
\usepackage{hyperref}
\usepackage{longtable}
\usepackage{enumitem}
\usepackage{lscape}

\begin{document}

\title{Facilitating Trustworthy Human-Agent Collaboration in LLM-based Multi-Agent System oriented Software Engineering}

\author{Krishna Ronanki}
\email{krishna.ronanki@gu.se}
\orcid{0009-0001-8242-6771}
\affiliation{%
  \institution{Chalmers University of Technology | University of Gothenburg}
  \streetaddress{Hörselgången 5}
  \city{Gothenburg}
  \country{Sweden}
  \postcode{417 56}
}

\renewcommand{\shortauthors}{Ronanki et al.}
\acmBooktitle{Companion Proceedings of the 33rd ACM Symposium on the Foundations of Software Engineering (FSE '25), June 23--27, 2025, Trondheim, Norway}

\begin{abstract}
    Multi-agent autonomous systems (MAS) are better at addressing challenges that spans across multiple domains than singular autonomous agents. This holds true within the field of software engineering (SE) as well. The state-of-the-art research on MAS within SE focuses on integrating LLMs at the core of autonomous agents to create LLM-based multi-agent autonomous (LMA) systems. However, the introduction of LMA systems into SE brings a plethora of challenges. One of the major challenges is the strategic allocation of tasks between humans and the LMA system in a trustworthy manner. To address this challenge, a RACI-based framework is proposed in this work in progress article, along with implementation guidelines and an example implementation of the framework. The proposed framework can facilitate efficient collaboration, ensure accountability, and mitigate potential risks associated with LLM-driven automation while aligning with the Trustworthy AI guidelines. The future steps for this work delineating the planned empirical validation method are also presented.
\end{abstract}

\keywords{Large Language Models, LLM-Based Multi-Agent Systems, Software Engineering, Trustworthy AI, Human-Agent Collaboration, DevOps}

\maketitle

\section{Introduction}

Software Engineering (SE) is an inherently collaborative field~\cite{rodriguez2021perceived} and SE collaboration is often artefact-based~\cite{4221622}. The advantage of SE collaboration being an artefact-based process is that the artefact evaluation against the ground truth can be automated~\cite{10449667}. This, alongside the observed fact that software engineers are increasingly becoming more reliant on the support of automation with increasing complexity in software systems~\cite{10.1145/3411764.3445650}, gave rise to the field of automated software engineering (ASE).

ASE focuses on developing methods and tools to assist humans and improve productivity in various tasks such as requirements traceability management, design specification, test data generation, defect tracking, and cost estimation with a focus on generation, storage and management of task specific artefacts~\cite{9772253}. The field of ASE has evolved from the application of traditional rule based tools such as static code analysers (SonarQube and ESLint), test automation tools (JUnit and pytest), dependency management tools (Maven and Gradle), and CI/CD tools (Jenkins and GitHub Actions) that automate a specific SE task to the development of frameworks that employ large language model(LLM)-based multi-agent autonomous (LMA) systems to automate multiple tasks within the software development lifecyle (SDLC) (ChatDev~\cite{qian-etal-2024-chatdev}, MetaGPT~\cite{hong2023metagpt}, and CAMEL~\cite{NEURIPS2023_a3621ee9}).

However, the introduction of general purpose foundation models such as LLMs into ASE brings a plethora of challenges. One of the major challenge is ensuring the implementation of LMA oriented SE aligns with the trustworthy AI guidelines and principles prescribed for the users of these technologies as per the EU AI Act (AIA)~\cite{AIAS}. Trustworthiness has been a issue for ASE even before the advent of LMA oriented SE. Multiple studies have shown that trust plays a critical and influential role within ASE~\cite{10.1007/978-3-319-07227-2_45,6112738,10.1145/2970276.2970347}.

Nonetheless, anthropomorphic features like natural language interaction and human-like reasoning enhance user trust in autonomous systems~\cite{WAYTZ2014113}. LLMs, which are capable of maintain context throughout conversations, excel in these areas~\cite{10062688}. Structured methods like chain-of-thought (CoT) prompting can further enhance their reasoning abilities~\cite{NEURIPS2022_9d560961}.

Even as the introduction of LMA systems within ASE raises the issue of trust, the unique characteristics of LLMs, if leveraged effectively, can be used as a means to increase the users' trust within the system. This however, does not solve all the pertinent issues of LMA oriented SE. 

One of the most pressing challenges in realising LMA oriented SE is facilitating trustworthy human-agent collaboration. Specifically, the strategic allocation of tasks between humans and the LMA system to leverage their unique individual strengths in an efficient and trustworthy manner~\cite{10.1145/3712003}.

To addressed this challenge that hinders human-agent collaboration in LMA oriented SE, a RACI-based framework along with implementation guidelines were developed to help define the roles and responsibilities of the LMA system and humans within a SDLC. An example implementation of the proposed framework is also provided by specifying a RACI matrix for a hypothetical implementation of LMA oriented SE within a DevOps environment. We chose the DevOps framework for the example implementation as it offers a structured SDLC with standardised processes and roles, which is observed to maximise the competences of the resources involved by decreasing manual work and offer more innovation capabilities~\cite{https://doi.org/10.1002/spe.3096}, all of which are crucial for the realisation of LMA oriented SE.

Section~\ref{section:background} provides relevant background and related works that contributed in the problem formulation and developing the solution space. Sections~\ref{section:method} present the framework and its implementation guidelines, and the example implementation of the framework in a DevOps environment. The article concludes within Section~\ref{section:conclusion} by outlining how this work in progress will be continued further. 

\section{Background} \label{section:background}

In this section, a brief overview of the concepts and state-of-the-art literature relevant to this article along with with nature of mentioned literature's relevance to this study are presented.

\subsection{LLM-based Multi-Agent (LMA) Systems}

When adopting an agent-oriented view of the world, it soon becomes apparent that most problems require or involve multiple agents~\cite{10.1007/3-540-48437-X_1}. This hold true for LLM-based agents as well. LMA systems are better capable of addressing real world challenges that are often spread across multiple domains and require expertise from different areas compared to singular LLM-based agents~\cite{10.1145/3712003}, including within SE~\cite{hong2023metagpt}. The LMA system leverages the capabilities of multiple specialised agents, each with distinct skills and roles. These agents collaborate seamlessly, working together to plan and execute tasks to achieve a shared objective~\cite{10.1145/3712003}.

\subsection{Trustworthy AI Requirements}

The Ethics Guidelines developed by European Commission's (EC) High-level expert group on artificial intelligence (AI HLEG) have identified four ethical principles that form the basis for trustworthy AI which include \textit{Respect for Human Autonomy}, \textit{Prevention of Harm}, \textit{Fairness}, and \textit{Explicability}~\cite{aihleg,AIAF}. But in order to achieve trustworthty AI in practice, they believe that there are seven key requirements that must be continuously assessed and managed throughout the entire lifecycle of an AI system:\textit{ Human Agency and Oversight}, \textit{Technical Robustness and Safety}, \textit{Privacy and Data Governance}, \textit{Transparency}, \textit{Diversity, Non-discrimination, and Fairness}, \textit{Societal and Environmental Well-being}, and \textit{Accountability}. The Trustworthy AI requirements are used as a reference within the proposed framework to ensure the human-LMA system collaboration is trustworthy.

\subsection{RACI Matrix}

The RACI matrix (Responsible, Accountable, Consulted, Informed) is a framework for defining roles and responsibilities within processes. It clarifies who is responsible for performing tasks (R), who is accountable for the outcome (A), who needs to be consulted to provide input (C), and who needs to be informed (I)~\cite{project2000guide}. In a LMA system, the RACI matrix can help in assigning clear roles between humans and the LLM-agents.

\subsection{Related Work}

He et al.~\cite{10.1145/3712003} conducted a systematic review of recent primary studies to map the current landscape of LMA applications across various stages of the software development lifecycle (SDLC). They identify critical research gaps and propose a comprehensive research agenda focusing on, among other things, optimising agent synergy and trustworthiness within an LMA system. 

Lin et al.~\cite{lin2024} introduce LCG, a LLM-based code generation framework inspired by established SE practices like waterfall, scrum, and test driven development(TDD). LCG uses multiple LLM agents to emulate software process models with roles such as requirement engineer, developer, and tester to collaboratively enhance code quality. Their results underscore the importance of adopting software process models to bolster the quality and consistency of LLM-generated code. This work inspired us to select the DevOps framework within the example implementation of the framework.

Zhang et al.~\cite{10.1007/978-3-031-61154-4_8} developed an Autonomous LLM-based Agent System (ALAS) and evaluated its impact on user story quality. As part of their experimental study, they created two LLM-agent profiles: agent product owner (PO), which is responsible for managing product backlog and prioritising user stories based on business value and customer needs and agent requirements engineer (RE), which concentrates on the quality of user stories. The agent profiles were designed to reflect the actual functions of POs and REs in agile teams. We based the characteristics of the predefined heterogenous LLM-agents~\cite{hong2023metagpt, 10.1145/3712003} within the example implementation of the framework based on agent PO and agent RE.

\section{Responsibility Assignment Framework} \label{section:method}

In this section, the proposed framework is presented for assigning responsibilities for the humans and the LLM-based agents of the LMA system. The definitions of each of the responsibility assignments are adapted to LMA oriented SE based on the original definitions of each of the four responsibility assignments of the standard RACI matrix.

\begin{itemize}
    \item \textbf{R-Responsible}: The actor(s) assigned this responsibility do the work to complete the task.
    \item \textbf{A-Accountable}: The actor(s) assigned this responsibility delegate the task to the \textit{responsible} actors and are responsible for the validation of the outcomes before the task is marked as complete.
    \item \textbf{C-Consulted}: The actor(s) assigned this responsibility provides inputs to guide the \textit{responsible} actor. The \textit{responsible} actor and the \textit{accountable} actor, if human, exercise their best judgement to decide how to use the \textit{consulted} actors' inputs. If the \textit{responsible} actor is an LLM-agent, then it should take the \textit{consulted} actor's input into account.
    \item \textbf{I-Informed}: The actor(s) assigned this responsibility need to be kept in the loop of the task execution without needing any explicit actions from the actor's end for the execution of the task.
\end{itemize}

\subsection{Framework Implementation Guidelines}

\begin{figure*} [ht!]
    \centering
    \includegraphics[width=0.9\linewidth]{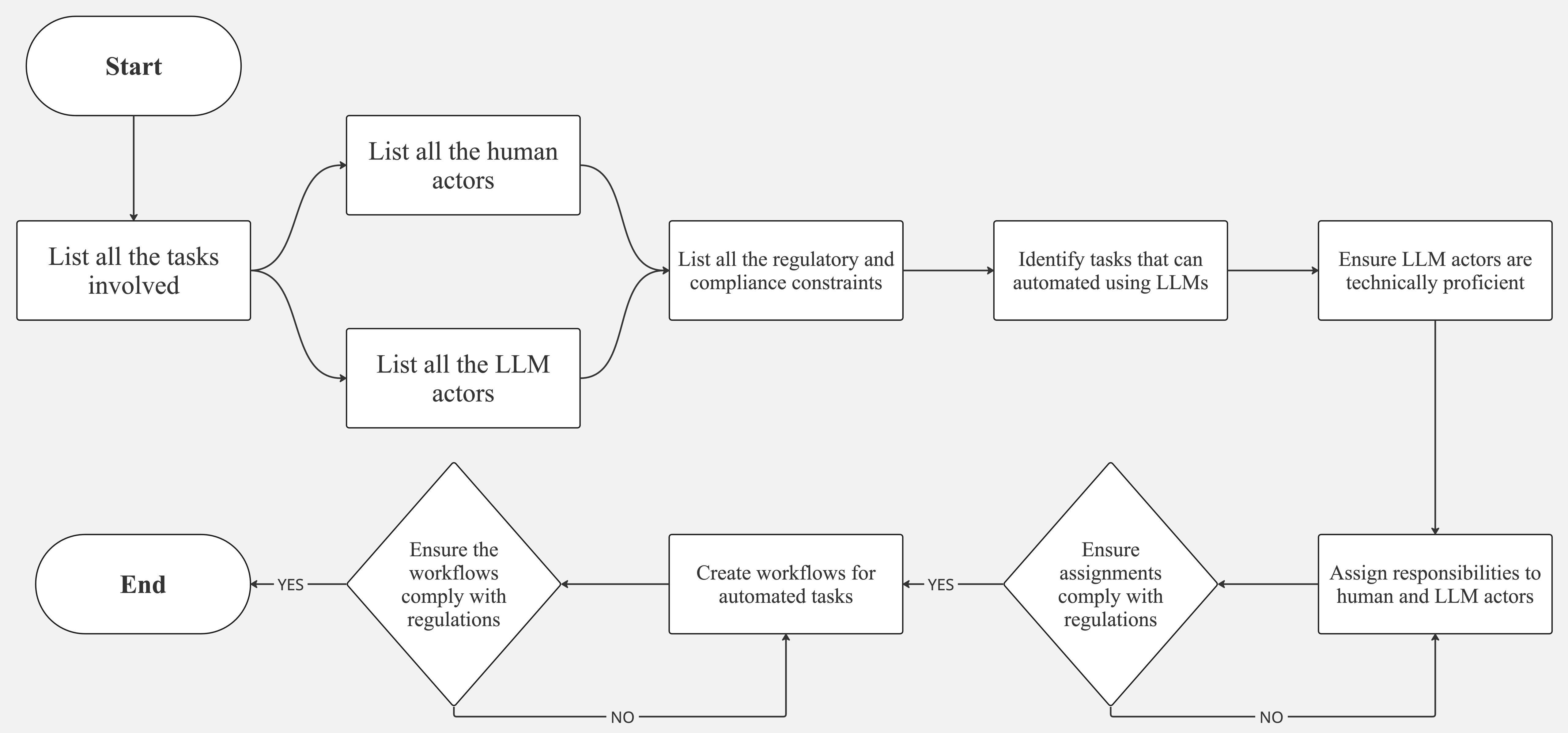}
    \caption{Responsibility Assignment Framework}
    \label{fig:fig-1}
\end{figure*}

\begin{itemize}
    \item \textbf{Step-1}: List all the tasks involved in the process where the LMA system will be implemented. Ensure all the tasks are artefact-based. 
    \item \textbf{Step-2}: List all the human actor(s) and LLM-agent(s) required for the completion of the task.
    \item \textbf{Step-3}: List all the applicable regulatory and compliance constraints on the actors involved.
    \item \textbf{Step-4}: Identify the tasks that can be automated using LLM-agents.
    \item \textbf{Step-5}: Ensure the LLM-agent(s) selected are technically proficient in automating the identified task(s).
    \item \textbf{Step-6}: Assign each actor across the row an R, A, C, or I to indicate the role they’ll play within the workflow to mark them as complete.
    \item \textbf{Step-7}: Ensure the assignments from \textbf{Step-6} are in do not conflict with the regulatory and compliance constraints. If yes, return to step \textbf{Step-5} and reassign the responsibilities to eliminate conflicts.
    \item \textbf{Step-8}: Create workflows for executing the automated tasks based on the responsibility assignments.
    \item \textbf{Step-9}: Ensure the created workflows do not conflict with the regulatory and compliance constraints. If yes, return to \textbf{Step-8} and redesign the workflows to eliminate conflicts.
\end{itemize}

This framework can be implemented to assign responsibilities for the human and LLM-agent actors for a single task or a suite of tasks within a SDLC phase or even an entire SDLC.

\subsubsection{Constraints of the Framework}

\begin{enumerate}
    \item There should be at least one `\textit{Responsible}' assignment for each task.
    \item There should be at least one `\textit{Accountable}' assignment for each task. The only exception is when there's no influence of the LMA system on the output of the task, i.e, the assignment of LLM-agent(s) is either '\textit{Informed}' or no assignment and the assignment of human(s) is `\textit{Responsible}' and/or '\textit{Accountable}'.
    \item There should be no `\textit{Responsible}' assignment for the LMA system or any of its agents if there's no '\textit{Accountable}' assignment for at least one human actor within a task.
\end{enumerate}

\subsection{Example Implementation}

In this example implementation of the proposed framework, the \textbf{Planning} phase of the DevOps SDLC is considered. Table~\ref{tab:devops_sdlc} provides the responsibility assignments for the human(s) and LLM-based agents for the collaborative execution of all the tasks within the planning phase.

\begin{itemize}
    \item \textbf{Step-1}: \textit{List all the tasks involved in the planning phase of the DevOps SDLC.} A typical planning phase within the DevOps SDLC has six tasks: requirements elicitation, creating product roadmap, creating features and user stories, creating product backlog, performing sprint planning and task allocation.
    \item \textbf{Step-2}: \textit{List all the human actor(s) and LLM-agent(s) required for the completion of the task.} In this scenario, we consider three human actors: product owner, business analyst, \& scrum master and three LLM-agents: LLM-agent A modelled after agent PO, LLM-agent B modelled after agent RE and LLM-agent C, a scrum implementation assistant, i.e, an LLM that has been fine-tuned for helping users implement scrum in practice\footnote{LLM-agent C is hypothetical agent that does not exist as of yet to the best of authors' knowledge. The existence of such an agent is an assumption made by the authors to provide the example implementation of the proposed framework}.
    \item \textbf{Step-3}: \textit{List all the applicable regulatory and compliance constraints on the actors involved.} Since the agents are based on LLMs, referring to the AIA, which is the most comprehensive in terms of coverage of applicable AI principles~\cite{10.1145/3597512.3599697}, is beneficial. If the LLM-agents are developed in-house from scratch, then the AIA's requirements for high-risk AI systems and GPAI requirements are applicable. If the LLM-agents include a third party developed models such OpenAI's GPT models for example, then the AIA's requirements for deployers of high-risk AI systems and GPAI models are applicable~\cite{AIAF}.
     \item \textbf{Step-4}: \textit{Identify the tasks that can be automated using LLM-agents.}\begin{enumerate}
         \item Since the requirements for a software system should come from the customer, the requirements elicitation task should not be automated. However, they can use the LLM-agents' assistance with the requirements elicitation. 
         \item Creating a product roadmap is a collaborative process involving multiple stakeholders to ensure the alignment of the product vision, establish collaboration, and aid risk management, which makes it not suitable to be completely automated using LLM-agents.
         \item Creating features and user stories is one of the tasks is suitable for automating using LLM-agent B, which is based on agent RE.
         \item Creating product backlog is next the task that is suitable for automating using LLM-agent A, which is based on agent PO.
         \item Sprint planning and task allocation, which is are typically performed by the scrum master are also suitable as they are artefact-based processes.
     \end{enumerate} 
    \item \textbf{Step-5}: \textit{Ensure the LLM-agent(s) selected are technically proficient in automating the identified task(s).} Since LLM-agent B is based on agent RE, it is capable of creating features and user stories. Similarly, as LLM-agent A is based on agent PO, it has the technical proficiency and domain knowledge required to create a product backlog. Since, LLM-agent C is fine-tuned for helping users implement scrum in practice, it has the technical proficiency and domain knowledge required to perform sprint planning and task allocation.
    \item \textbf{Step-6}: \textit{Assign each actor across the row an R, A, C, or I to indicate the role they’ll play within the workflow to mark them as complete.} \begin{enumerate}
        \item For reasons described in \textbf{step-4}, the \textit{responsible} actor for requirements elicitation task is the business analyst and \textit{accountable} is the product owner. The LLM-agent B and LLM-agent C can be the \textit{consulted} actors to leverage their domain knowledge while the scrum master and LLM-agent A can be the \textit{informed actors} as keeping them in the loop will be helpful with providing them with the overall context of LMA system being implemented.
        \item Similarly, for creating the product roadmap task, the \textit{responsible} actor is the product owner. The LLM-agent A and LLM-agent C are \textit{consulted} actors while the scrum master and business analyst are the \textit{informed} actors. There is no \textit{accountable} assignment in this scenario as per the second framework constraint. The \textit{responsible} actor is the \textit{accountable} actor as well.
        \item For creating features and user stories, as described in \textbf{step-4}, the \textit{responsible} actor will be LLM-agent B while the business analyst who will use the LLM to generate the user stories and features will be the \textit{accountable} actor. LLM-agent A can be helpful in this scenario so it can be the \textit{consulted} actor and rest can be either \textit{informed} or no assignment based on the necessity.
        \item Similarly, for creating the product backlog task, the \textit{responsible} actor will be LLM-agent A while the product owner who will use the LLM will the \textit{accountable} actor. As it is valuable to keep the business analyst and scrum master in loop to enhance the collaboration, they can be either \textit{consulted} or \textit{informed}, based on the necessity while the rest of the actors can remain unassigned for this task. 
        \item For sprint planning and task allocation, the scrum master who is also the \textit{accountable} actor, can use LLM-agent C, which is the \textit{responsible} actor, to generate the required outputs to complete the task. The product owner can be the \textit{informed} actor to keep them in the loop.
    \end{enumerate} 
    \item \textbf{Step-7}: \textit{Ensure the assignments from \textbf{Step-6} are in do not conflict with the regulatory and compliance constraints. If yes, return to step \textbf{Step-5} and reassign the responsibilities to eliminate conflicts.} The AIA's obligations and requirements do not apply to the tasks where LLM's responsibility is just informed. Since there is a human actor who is assigned the \textit{accountable} responsibility whenever there is either a \textit{responsible} or an \textit{informed} responsibility for the LLM-agent, there is no conflict with the AIA's obligations and requirements.
     \item \textbf{Step-8}: \textit{Create workflows for executing the automated tasks based on the responsibility assignments.} \begin{enumerate} 
        \item Although the requirements elicitation and creating product roadmap tasks have been identified as not suitable for automation, LLM-agents can still be leveraged to assist the human actors in these tasks. For example, the business analysts within the requirements elicitation task can use the LLM-agent B and LLM-agent C to create the requirements elicitation questionnaire or planning the elicitation meetings. Similarly, the product owner in the creating product roadmap task can leverage LLM-agent A and LLM-agent C's assistance.
        \item For the remaining four tasks, the workflows can be designed in such a way that the final output of the user interaction with the LLM-agents will yield a verifiable artefact. For example, the features \& user stories will be generated using the LLM-agent based on the requirements elicited by the business analyst from the clients. The business analyst needs to verify the outputs before the product backlog can be generated by the LLM-agent based on the list of features and user stories generated. This time, the product owner needs to verify the product backlog before the sprint planning and task allocation can be performed. The outputs of the LLM-based sprint planning and taks allocation needs to verified by the scrum master before they can implemented in practice. 
    \end{enumerate}
    \item \textbf{Step-9}: \textit{Ensure the created workflows do not conflict with the regulatory constraints. If yes, return to \textbf{Step-8} and redesign the workflows to eliminate conflicts.} Since, the first two tasks don't involve the use of LLMs in the workflow, the AIA's  requirements do not apply. For the next four tasks, since the technical capabilities of the LLM-agents has been ensured within \textbf{step-5} and human oversight and validation mechanisms were ensured within \textbf{step-8}, they do not conflict with the AIA's requirements for deployers.
\end{itemize} 

\begin{table*}[ht!]
    \centering
    \resizebox{0.95\linewidth}{!}{
    \begin{tabular}{|l|l|l|l|l|l|l|l|}
    \hline
    Task & Product Owner & Business Analyst & Scrum Master & LLM-agent A & LLM-agent B & LLM-agent C\\
     \hline
    Requirements elicitation & A & R  & I & I & C & C\\
     \hline
    Create product roadmap & R & I & I & C & - & C\\
     \hline
    Create features \& user stories & I & A & I  & C & R & -\\
     \hline
    Create product backlog & A & I/C &  I &  R & - & -\\
     \hline
    Sprint planning & I & -  &  A &  C & - & R\\
     \hline
    Task allocation & I & - &  A & C & - & R\\
     \hline
    \end{tabular}}
    \caption{RACI matrix for the example implementation of the proposed framework}
    \label{tab:devops_sdlc}
\vspace{-10pt}
\end{table*}

The RACI matrix for the example implementation of the proposed framework based on above description is provided in Table~\ref{tab:devops_sdlc}. This is, however, an example implementation of the proposed framework using a hypothetical example. The framework allows practitioners to customise and adapt the output of RACI matrix based on the the SDLC framework and the standard operating procedures (SOP) specific to their organisation in a flexible manner.

\section{Conclusion and Future Work} \label{section:conclusion}

The proposed framework leverages the RACI matrix to define the roles and responsibilities of humans and LLM-based agents in the LMA system. This structured role delineation enhances collaboration, ensures accountability, and mitigates potential risks associated with LLM-driven automation. By systematically distributing decision-making authority and oversight, the framework can aid in optimising the efficiency of the LMA system based SE workflows while maintaining human control over critical software development activities, aligning with the Trustworthy AI guidelines.

Despite its theoretical promise, the framework has not yet undergone empirical validation. Moving forward, a groupware walkthrough~\cite{10.1145/503376.503458} based multi case study will be conducted to collect empirical validation data from \textbf{experts/potential users}. 

\bibliographystyle{ACM-Reference-Format}
\bibliography{reference}
\balance

\end{document}